# Triplicity and Physical Characteristics of Asteroid (216) Kleopatra

P. Descamps[1], F. Marchis[1,2,11], J. Berthier[1], J. P. Emery[3], G. Duchêne[2,4], I. de Pater[2], M. H. Wong[2], L. Lim[3], H.B. Hammel[5], F. Vachier[1], P.Wiggins[6], J..-P. Teng-Chuen-Yu[7], A. Peyrot[7], J. Pollock[8], M. Assafin[9], R. Vieira-Martins[10,1], J.I.B. Camargo[9,10], F.Braga-Ribas[9,10], B. Macomber[11]

[1] Institut de Mécanique Céleste et de Calcul des Éphémérides, Observatoire de Paris, UMR8028 CNRS, 77 av. Denfert-Rochereau 75014 Paris, France
[2] University of California at Berkeley, Department of Astronomy, 601 Campbell Hall, Berkeley, CA 94720, USA
[3] University of Tennessee at Knoxville, 306 EPS Building, 1412 Circle Drive, Knoxville, TN 37996, USA
[4] Université Joseph Fourier – Grenoble 1/CNRS, Laboratoire d'Astrophysique de Grenoble (LAOG), UMR5571, BP 53, 38041 Grenoble Cedex 09, France
[5] Space Science Institute, Boulder, CO 80303, USA
[6] Tooele Utah 84074, USA
[7] Makes Observatory, 18, rue G. Bizet - Les Makes - 97421 La Rivière, France
[8] Appalachian State University, Department of Physics and Astronomy, 231 CAP Building, Boone, NC 28608, USA
[9] Observatório do Valongo/UFRJ - Universidade Federal do Rio de Janeiro, Rua Ladeira Pedro Antonio, 43 CEP 20080-090 Rio de Janeiro, Brazil
[10] Observatório Nacional/MCT, R. Gal. José Cristino 77, CEP20921-400, RJ, Brasil.
[11] SETI Institute, Carl Sagan Center, 515 N. Whisman Road, Mountain View CA 94043, USA

Pages: 35

Tables: 3

Figures: 9

*Corresponding author:*
**Pascal Descamps**
IMCCE, Paris Observatory
77, avenue Denfert-Rochereau
75014 Paris
France

descamps@imcce.fr
Phone: +33 (0)140512268
Fax:    +33 (0)146332834

**Abstract**

To take full advantage of the September 2008 opposition passage of the M-type asteroid (216) Kleopatra, we have used near-infrared adaptive optics (AO) imaging with the W.M. Keck II telescope to capture unprecedented high resolution images of this unusual asteroid. Our AO observations with the W.M. Keck II telescope, combined with Spitzer/IRS spectroscopic observations and past stellar occultations, confirm the value of its IRAS radiometric radius of 67.5 km as well as its dog-bone shape suggested by earlier radar observations. Our Keck AO observations revealed the presence of two small satellites in orbit about Kleopatra (see Marchis et al., 2008). Accurate measurements of the satellite orbits over a full month enabled us to determine the total mass of the system to be $4.64 \pm 0.02 \; 10^{18}$ Kg. This translates into a bulk density of $3.6 \pm 0.4$ g/cm$^3$, which implies a macroscopic porosity for Kleopatra of ~ 30-50%, typical of a rubble-pile asteroid. From these physical characteristics we measured its specific angular momentum, very close to that of a spinning equilibrium dumbbell.

# 1. Introduction

(216) Kleopatra is one of the most intriguing Main-Belt asteroids. It is classified as an M-type asteroid in the Tholen et al. (1989) taxonomy or Xe-type in the Bus and Binzel (2002) taxonomy because of its featureless spectrum in the visible and near-infrared. It has an IRAS radius of 67.5±1.1 km and an associated visual albedo $p_v$ = 0.1164 (Tedesco et al., 2002). This asteroid has received a lot of attention over the last 30 years because of its significant brightness variations (up to 1.2 mag) over a short period of time ($P_{spin}$ = 5.385 h). Early on these lightcurve amplitudes with values depending on the relative positions of (216) Kleopatra and the Earth suggested that the shape of (216) Kleopatra might be elongated or bilobed (Scaltriti and Zappala, 1978, Tholen, 1980, Zappalà et al., 1983). After the stellar occultation of January 1991 it was described as cigar-shaped (Dunham et al., 1991). Different models were later proposed for Roche equilibrium ellipsoids (Cellino et al., 1985) or contact binary (Tanga et al., 2001). AO imaging in 1999 with the ESO's 3.6 meter telescope on La Silla showed Kleopatra as a possible nearly-contact binary system (Marchis et al., 1999, Hestroffer et al., 2002b) confirming the bimodality of the wide-bandwidth spectra highlighted in previous radar observations (Mitchell et al., 1995). Ostro et al. (2000) performed comprehensive radar observations and inferred a three-dimensional reconstructed shape similar to a dog-bone, with dimensions of 217x94x81km and an equivalent radius of 54.25 ± 7km (radius of a sphere with the same volume as the shape model), which has been widely used ever since.

Since then this dog-bone model has been extensively tested using a variety of data from ground-based photometry, stellar occultations, high resolution interferometric data from the HST/FGS (Hestroffer et al., 2002a) and polarimetric observations (Takahashi et al. 2004). Hestroffer et al (2002) suggested that (216) Kleopatra most likely is very elongated, as suggested before; however, the authors could not determine whether the asteroid did consist of two components separated by a small gap, or whether it was indeed one elongated object. Takahashi et al. (2004), from lightcurve simulations, tend to favor the bilobed model of Tanga et al. (2001). These works presented serious caveats and evidence showing that the dog-bone shape model needs to be somehow revised and improved or at the very least compared with new imaging data with even higher spatial resolution.

Besides its puzzling shape, asteroid (216) Kleopatra was also a target of interest for those searching for satellites orbiting minor planets. The discovery of companions of main belt asteroids brings a wealth of information about the asteroid and its origin. In 1993, Storrs et al. (1999) searched for satellites around 10 asteroids including (216) Kleopatra using WF/PC on the Hubble Space telescope. They did not find any companion around but at that time, the asteroid was very far from Earth at a distance of 2.38 AU. In addition, these observations were performed with the first generation of HST instruments so data were significantly poorer in contrast and angular resolution that can be achieved today with Adaptive Optics (AO) systems on large ground-based telescopes. We note that that they did not report the detection of a satellite around the Trojan asteroid 624 Hektor whereas Marchis et al. (2006) detected a 5km-satellite around it using the Keck AO system in 2006. We therefore decided to conduct another search for satellites orbiting (216) Kleopatra during its particularly favorable 2008 opposition when its geocentric distance was only 1.23 AU, half as close as before.

The 2008 opposition provided an important opportunity to obtain images of (216) Kleopatra at high angular resolution and to search for moonlets around the object since the next similar configuration will not occur again until October 2013. In this work we present results of high-resolution imaging observations of (216) Kleopatra made with the W. M. Keck II telescope in late 2008.

In section 2 we report on the discovery of two small satellites orbiting (216) Kleopatra. In addition to searching for satellites around (216) Kleopatra, our observations actually resolved the disk of the primary target. These resolved images are presented in section 3. They provide, coupled with thermal infrared spectroscopy and revisited past stellar occultations, new direct and conclusive information on the size and shape of (216) Kleopatra's primary that is required in determining its bulk density. Lastly, the measurement of the period and semi-major axis of the companions' orbits allowed the determination of the mass, the bulk density and the specific angular momentum of the system (section 4).

## 2. Discovery of two satellites using Keck adaptive optics imaging

(216) Kleopatra was imaged in moderate angular resolution only once, in 1999, with the ESO 3.6m telescope and its AO system called ADONIS (Marchis et al., 1999, Hestroffer et al., 2002b). However the relatively large image scale (35 milliarcseconds/pixel) combined with the modest spatial resolution (0.123" in Ks band) severely limited the image quality.

In September 2008, (216) Kleopatra approached to a geocentric distance of 1.23 AU, similar to the conditions of November 1999, with a subtended angular size of ~0.20". AO observations were carried out in September-October 2008 on the 10-m Keck telescope. With a pixel size on the sky of 9.942±0.050 milliarcseconds (*mas*) offered on the near-infrared camera NIRC2 and an expected FWHM of 0.034" in FeII ($\lambda_c$ = 1.6455 µm, $\Delta\lambda$=0.0256 µm), 0.025" in Jcont ($\lambda_c$ = 1.2132 µm, $\Delta\lambda$=0.0198 µm), 0.033" in H ($\lambda_c$ = 1.633 µm, $\Delta\lambda$=0.0296 µm) and 0.047" in Kcont ($\lambda_c$ = 2.2706 µm, $\Delta\lambda$=0.0296 µm), we collected detailed images of (216) Kleopatra with a spatial resolution four times better than previous AO observations. These exquisite images revealed the existence of two small satellites, making (216) Kleopatra the fourth triple asteroid discovered among the main belt asteroids (Marchis et al., 2008b) after 87 Sylvia (Marchis et al., 2005), 45 Eugenia (Marchis et al., 2007) and 3749 Balam (Marchis et al., 2008a). For the sake of clarity we adopt from now on the designation *(216)*

*Kleopatra* for the whole system and *Kleopatra* for the primary alone. Kleopatra's companions are 6 magnitudes fainter than their primary body and are separated by about 0.50 and 0.76 arcsec respectively in September 2008 (Fig. 1).

Shortly after this discovery, we collected additional observations of (216) Kleopatra to be able to estimate the orbital parameters of the satellites while taking advantage of the close proximity of the triple system from Earth. In Table 1 we report the measured relative positions of satellites with respect to the center-of-light of Kleopatra from observations collected in September and October 2008. These astrometric positions were derived after fitting the moonlet and primary centroid by a Moffat-Gauss 2D stellar profile from a specific method suited to adaptive optics images (Descamps et al., 2002). The 1-sigma error on the position of the satellites is estimated to 10 *mas*.

A preliminary orbit solution was derived assuming a simple Keplerian formalism (Descamps, 2005). The resulting orbital elements are listed in Table 2. Figure 2 shows the residuals between observed and computed positions of each satellite. The root mean square (RMS) error on the whole set of data is 30*mas* for the outermost satellite and 15*mas* for the innermost, resulting in less than a 3-sigma error in astrometric position. We determined the Keplerian periods after adopting a dynamical flattening $J_2$ of 0.6, derived from the 3D polygonal dog-bone shape model (Ostro et al., 2000). The orbit pole directions are within a few degrees of the pole solution of the primary found by Ostro et al. (2000) which was given by its ecliptic coordinates $\lambda = 72°$ and $\beta = +27°$ ECJ2000.

Eventually we adopted an average solution for the primary pole based on the assumption that the equatorial and orbital planes of the satellites are parallel. Our adopted primary spin pole is given in J2000 ecliptic coordinates by $\lambda = 76 \pm 3°$ and $\beta = +16 \pm 1°$. Based on peak-to-peak differential amplitude between each secondary with the primary, the sizes of the outer and inner satellites were estimated to 8.9±1.6 km and 6.9±1.6 km respectively using the IRAS diameter for Kleopatra itself.

## 3. New insights on the shape and size of Kleopatra

### 3.1 Ground-based photometry

During the past 3 years, we performed extensive photometric observations of Kleopatra while it was viewed nearly pole-on in 2007 and edge-on in 2008. Observations were acquired using small reflector telescopes since the target magnitude in V was typically in the range 12-13. At the Makes Observatory (La Réunion Island, observatory code 181), we used a 0.35m (C14) with a CCD SBIG ST-7 camera (765 x 510 pixels) in V filter. Observations at LNA (Laboratório Nacional de Astrofísica, Brazil, observatory code 874) were carried out with the 0.6m Boller & Chivens telescope using an R filter. Observations at Lick Observatory were recorded using the Nickel -1m telescope and its CCD#2 camera in R band. The robotic 0.6 m telescope super-LOTIS located at Kitt Peak was also used to record R band observations. Lastly, some observations were made with the 16-inch telescope (0.40m) of the Rankin Science Observatory (RSO) at Appalachian State University. Amateur astronomer P. Wiggins in Tooele, Utah, significantly contributed with a 0.35 m telescope. Our group also used the 1.20m telescope of Haute-Provence observatory to collect photometric data of Kleopatra.

A few lightcurves are displayed in Figure 3. Kleopatra's visual lightcurves have amplitudes as large as 1.4 magnitudes. Two distinct maxima and two distinct minima clearly dominated by the signal of the large primary (synodic spin period of 5.385 h). A non-axisymmetric shape for the primary is also evident from the unequal maxima and unequal minima in Kleopatra's lightcurves.

In order to compare photometric observations with predicted synthetic lightcurves we used the 3D polyhedron radar dog-bone shape model. Each vertex is given by its x y z coordinates. With this technique, any asteroid shape can be represented by a set of facets. Once the line of sight and the direction of Sun are known, it suffices to select the facets that are both visible by the observer and illuminated by the Sun. The total reflected light is then computed by adding the contribution of each of these *active* facets.

In the special case of a concave shape model, self-shadowing occurs at non-zero phase angles. This effect must be taken into consideration because it represents a non-negligible amount of the overall brightness variation. Each facet reflects the solar light according to its orientation with respect to Earth and Sun as well as the adopted scattering law. We adopted the five parameters Hapke law (1981). Hapke parameters are derived from the conversion equations proposed by Verbiscer and Veverka (1995) which express each Hapke parameter as a function of the slope parameter $G$ and the geometric albedo $p_v$. We used $G = 0.26 \pm 0.07$ (Lagerkvist and Magnusson, 1990), $p_v = 0.1164$ (inferred from IRAS observations) and a typical macroscopic roughness $\theta = 20°$ (Helfenstein and Veverka, 1989). We inferred a single scattering albedo $\omega = 0.181$, a regolith compaction parameter which characterizes the width of the opposition surge $h = 0.047$, the amplitude of the opposition surge $B_0 = 1.276$ and the Henyey-Greenstein asymmetry parameter $g = -0.254$.

The 3D-shape model fails to perfectly reproduce the observed lightcurves. This is particularly noticeable in the observations made in summer 2008. At that time Kleopatra's primary was seen edge-on with a high phase angle. Under this aspect, the unusual shape of Kleopatra (very long with two protuberances at each end) caused conspicuous shadows to be cast on the surface of the asteroid. This phenomenon explains the very high amplitude of 1.4 magnitudes found in the photometric variation.

The radar shape model fails to account satisfactorily for this phenomenon, especially as regards the asymmetry between the two minima of the lightcurve. Nonetheless, apart from ~0.3mag deviations between model and data for the highest amplitude troughs in this extreme configuration, the 3D radar shape model provides a very good overall match to the observed light curves. Consequently the radar shape model was adopted for the rest of the study despite its shortcomings.

**3.2 Direct imaging from adaptive optics**

Figure 4 shows views of Kleopatra imaged in the near infrared with the 10-m W.M. Keck telescope in September and October 2008 with a resolution of 35 *mas*. Kleopatra was not far from an edge-on aspect. For each of those images the aspect of the primary was computed using the dog-bone radar shape model and projected onto the plane of sky. In each of the images the asteroid's north is up. The contours provided by the shape solution of Kleopatra corresponding to its IRAS equivalent diameter, 135 km, and to the Ostro et al (2001) value of 108.5 km have been overlaid on the AO views. Optimal resolution was restored by applying the AIDA deconvolution algorithm (Hom et al. 2007). The silhouette of the large primary was extracted with a Laplacian filter. Surface brightness variations on the final image are artefacts resulting from the Laplacian filtering.

Kleopatra is confirmed to be irregularly shaped and strongly bifurcated. Its spatial extent is about 271x65 km. It appears to look more like a dumb-bell than a dog-bone. Two equal-sized misshapen lobes, each 80 km across, are joined by a thin and long bridge of matter about 50 to 65 km across and 90 km long. From a quick comparison between AO images and projected radar model, it appears that Kleopatra seems much narrower than its radar model and therefore is likely a more stretched body than believed. Despite these slight disagreements the radar shape model is a good reference to estimate the size of Kleopatra. From the contours which have been superimposed, it is undeniable that the one corresponding to the IRAS size best matches the real shape of the primary.

We have now sufficient material (AO images and lightcurves) to theoretically derive the non-convex shape solution from the AO+LC inversion method (Kaasalainen and Torppa 2001) as it was already done for 121 Hermione (Descamps et al., 2009a). However this method is presently restricted to "starlike" objects. A starlike figure is not strongly bifurcated and can have concavities, but not so extreme that the surface folds back on itself as seen from the center of mass. That is, the radius vector from the center to any point on the surface is single-valued. A "non-starlike" figure has concavities so extreme that a radius vector in certain directions is multi-valued. Therefore the radar shape model of Kleopatra was still kept to determine the equivalent radius of Kleopatra from infrared Spitzer observations.

### 3.3 Spitzer observation

(216) Kleopatra was a target in a Spitzer telescope program in Cycle-2 (PI: L. Lim, GO #20481). The aim of this program was to investigate the thermal properties and compositions of M-type asteroids by high S/N spectroscopic study in the thermal infrared. Spectra of (216) Kleopatra from 5.2 to 35 microns were measured using the InfraRed Spectrograph (IRS) instrument in both low (SL2 and SL1; 5.2 to 14.2 μm; R=64 to 128) and high (SH and LH; 9.9 to 35 μm; R=600) spectral resolution modes on February 2, 2006 from 06:07 to 06:21 UT. The combined processed thermal spectrum is shown in Fig. 5.

Thermal emission from an asteroid is dependent on the size (projected area) and temperature distribution. At this time, the asteroid was not known to be multiple. The surface areas of the moons of Kleopatra are negligible compared with the large primary. The IRAS data were used together with an assumed temperature distribution (based on a spherical shape) to estimate the size (Tedesco et al., 2002). The Spitzer telescope data offer an improvement in that the high sensitivity and broad wavelength coverage enable fitting of the temperature distribution as well as the size, resulting in an accurate estimate of albedo as well as size. This Spitzer spectrum was analyzed using a version of the NEATM model (Harris, 1998) that was modified to incorporate ellipsoidal shapes (e.g., Brown, 1985, Lim et al. 2010). The ellipsoidal version of NEATM assumes that the surface temperature varies as $\cos^{1/4}$ of the solar zenith angle and integrates the thermal flux from this temperature distribution over the ellipsoidal shape. The model incorporates a factor known as the beaming parameter ($\eta$) to adjust the temperature as a proxy for the effects of surface roughness and thermal inertia (Lebofsky and Spencer, 1989). The value of $\eta$ scales the subsolar temperature, where $\eta = 1$ corresponds to a smooth thermal-inertia surface, $\eta > 1.0$ means the temperature is lower than expected, suggesting a relatively high thermal inertia, whereas $\eta < 1.0$ indicates a higher temperature, suggesting a fairly rough surface.

From these fits to the Spitzer data, we derive an effective radius (i.e., the radius of a sphere with the same projected surface area) of $R_{eff} = 78.2 \pm 2.9$ km and $\eta = 1.25 \pm 0.03$. Combining this $R_{eff}$ with the absolute magnitude ($H_V = 7.30$) yields a visible geometric albedo ($p_v$) of $0.087 \pm 0.007$, somewhat lower than the estimate from IRAS. Using our pole solution and the 3D radar shape model of Kleopatra primary, we calculate the expected appearance and projected area as viewed by Spitzer (Fig. 5). This model calculation results in a model $R_{eff}$ of 78 km corresponding to an equivalent radius $R_{eq}$ of 67.5 km, in perfect agreement with the results from Spitzer/IRS. This also agrees well with the IRAS radius and with our own measurements from AO images (see section 3.2). Eventually, by retaining the uncertainty associated with the Spitzer determination, we adopt an equivalent radius of $67.5 \pm 2.9$ km.

### 3.4 Stellar occultation

Another highly accurate type of observation for constraining an asteroids size and shape is the recording of a stellar occultation. Eight occultations of a star by (216) Kleopatra were reported in the past (Dunham et al., 2009). Among them only two events (observed in 1980 and 1991) contained a sufficient number of chords to be usefully compared with our model.

### 3.4.1 The 1980 event

On 1980 October 10, the occultation of a bright star (HIP 114784, V = 8.85) by (216) Kleopatra was predicted to cross the Pacific Northwest part of the USA. At that time (216) Kleopatra was suspected to be a contact binary because of its unusually high lightcurve amplitude (see Sky and Telescope, Sept. 1980). This led to a special request being sent asking observers to go to the eastern track and to California. Unfortunately, photometric data showed that the occultation would occur near the minimum of light with Kleopatra facing nearly end-on and making it impossible to distinguish between a contact binary model or an elongated shape based on the observed occultation outline (Dunham, 1981). Nevertheless the observation of the occultation was a success. Fourteen stations monitored the event, eleven of which recorded a positive extinction of the star, including two recorded from northern California (Loma Prieta) about 475 km away from the path of Kleopatra's shadow on Earth. These observers, G. Rattley and B. Cooke, reported a $0^s.9$ duration event from sites 610 m apart. Since their site was located outside the path of the primary occultation, it was thought that their observation could be a possible secondary event. No such event was recorded at Lick Observatory which was only 3.4 km east of their position. Now, nearly 30 years later, we are finally in a position to say that they probably observed the occultation of the target star by one of Kleopatra's two satellites, as demonstrated below.

Figure 6 shows the observed chords (Dunham et al., 2009, and Dunham, private communication) including both actual occultation events and negative observations, as well as the lone "off-axis" short event. The number and the distribution of the chords along the cross-section of Kleopatra (Fig. 7) enable us to compare the observed chords and the shape contour of Kleopatra derived from the radar shape model adopting an equivalent radius of 67.5 km. It shows that the overall size and shape of the model are in good agreement with the chords despite some minor discrepancies. They can be explained by time errors made by observers, as well as by uncertainties in the theoretical shape model.

The single chord of the secondary event (Fig. 8) combined with the external negative chords (recorded at Lick observatory) allows us to estimate a diameter of 8±4 km for the occulting moonlet which is in fairly good agreement with our estimation from AO images (section 2 and Table 2). Its astrometric relative position with respect to the primary is measured as $\Delta\alpha cos\delta$ = 0".2711 ±0".0002, $\Delta\delta$ = -0".4564 ±0".0002. Taking into account the proximity of the orbit of the external satellite, we conclude that the occulting satellite is likely S/2008 ((216)) 1. This is supported by a new fit of its orbit parameters including this new astrometric position, while the fit has no solution for the second satellite. The separation between the predicted orbit and the observed chord is less than 30 km (40 mas), which is better than the accuracy of all the astrometric positions used to fit the orbit.

### 3.4.2 The 1991 event

On 1991 January 19, eleven observers spanned across the Northern USA recorded eight positive extinctions of another star occulted by (216) Kleopatra (Dunham et al., 1991). The corresponding chords are projected onto the sky plane in Fig.9 (reproduced from the abstract for the 1991 International Conference on Asteroids, Comets, Meteors). They sketched an outline about 55 km wide and 230 km long. The occultation occurred near the maximum of (216) Kleopatra's lightcurve at the central meridian longitude of 152° indicating that it was broadside to the observers. This was the first observation of Kleopatra in which the primary was described as cigar shaped. In two of our 2008 AO images the longitude of the sub-Earth central meridian was very close that during the occultation (see Table 3). The only difference between these two epochs is in the latitude of the sub-Earth point (30° from each other) and the position angle of the north pole. Nevertheless, after rotating the AO images to align the pole directions the outlines of Kleopatra were retrieved and projected onto the occultation plane (Fig. 9). The agreement between the projected profiles extracted from the 2008 AO images and the 1991 recorded chords is satisfactory.

**4. Bulk density and macroscopic porosity estimated from the measured mass of Kleopatra**

Measurements of asteroid bulk density, along with data on the grain densities of analogue meteorites, provide valuable insight into asteroid porosity and internal structure. Therefore the bulk density is one of the most important physical characteristics to be determined. So far, the bulk density of only one M-type binary asteroid, 22 Kalliope, has been measured (3.4 ± 0.2 g/cm$^3$) based on a combination of a 6-year campaign of AO observations and mutual events observations observed in spring 2007 (Descamps et al., 2008).

From a simple Roche model, Cellino et al. (1985) inferred a bulk density of Kleopatra of 3.9 g/cm$^3$. In the present work we are able to straightforwardly measure the total mass of the system using Kepler's third law based on the orbits of its satellites. It is estimated to be $M_{Total}$ = 4.64 ± 0.2 10$^{18}$ kg. Using the equivalent radius of Kleopatra of 67.5±2.9 km as derived in Section 3.3, we infer an average bulk density of $\rho$ = 3.6 ± 0.4 g/cm$^3$.

(216) Kleopatra is classified as a M-type asteroid. Given their spectral properties, the most likely meteoritic analogs for M-class asteroids and their associated grain densities are irons (~7.5 g/cm$^3$), CH/CB/bencubbinite (metal rich) carbonaceous chondrites or silicate-bearing iron meteorites (~5.0

g/cm$^3$), enstatite chondrites (~3.5 g/cm$^3$), and carbonaceous chondrites (~2.5 g/cm$^3$) (Shepard et al., 2008).

(216) Kleopatra's radar albedo of 0.7 (Mitchell et al. 1995; Ostro et al. 2000) exceeds those of all other radar detected Main-Belt asteroids (the next highest is 0.28 for 16 Psyche) and led to the interpretation of a mostly metallic composition. Moreover, its low circular polarization ratio ($\mu_c$= 0.00±0.05) distinguishes (216) Kleopatra as one of the smoothest radar-detected targets at centimeter-to-meter scales (Mitchell et al., 1995). Kleopatra also has the highest known thermal inertia for a Main-Belt asteroid, larger than 50 Jm$^{-2}$K$^{-1}$s$^{-1/2}$, consistent with a "metallic" surface (Mueller et al., 2005). Additionally, spectroscopic study by Rivkin et al. (1995) did not reveal any 3µm absorption features implying that (216) Kleopatra is not hydrated. Given the bulk density that we derived (3.6 ± 0.4 g/cm$^3$), the candidate meteoritic analogs should have a grain density no less than ~4 g/cm$^3$ which quite obviously rules out enstatite (typical density of 3.2-3.9 g/cm$^3$) and carbonaceous chondrites (density of ~2.3 g/cm$^3$) analogues. On the other hand, Hardersen et al. (2005), based on spectral results combined with its high radar albedo, suggested that Kleopatra is most probably exposed core fragments of differentiated bodies with a high metal abundance. They inferred from an empirical relationship between radar albedo and density (Ostro et al., 1985) an estimated surface density of 6.92 g/cm$^3$. Finally, these lines of evidence tend to favor a highly metallic surface. In that case, our bulk density of 3.6 g/cm$^3$ suggests a high macroscopic porosity in the range 30–50%, typical of extensively fractured or loosely packed asteroids (Britt et al., 2002).

## 5. Discussion

In a recent work, Descamps and Marchis (2008) set forth the idea that most of the Main-Belt binary systems with a large rubble-pile primary might have originated in a rotational fission process. Adopting their formalism, we can update, from the full set of physical parameters determined in the present work, the normalized spin rate[1] of Kleopatra to $\Omega = 0.318 \pm 0.045$ and the non-sphericity parameter[2] $\lambda=3.65$, which is the ratio between the moment of inertia of the body with respect to its spin axis and the moment of inertia of the equivalent sphere. The specific angular momentum of the system is then computed from a formula given in appendix A of the same work, $H = 0.47$. This places Kleopatra in the vicinity of the equilibrium dumbbell branch for homogeneous spinning liquid masses, well beyond the classical instability limit of the Jacobi ellipsoids $H = 0.39$ (Weidenschiling, 1981) which has long been accepted as an upper limit for a single body.

We can now draw a convincing picture of Kleopatra's story, born from the aftermath of a reaccumulation process following the violent disruption of a parent body and sped up fast enough to deform according to the rough guidelines given by the equilibrium sequences of a spinning liquid mass. Such a scenario was already invoked to account for the way the synchronous double system of 90 Antiope, endowed with the same total specific angular momentum, was formed (Descamps et al. 2007; Descamps et al. 2009b). This does not mean that the interiors are liquid but only that the loosely packed internal structure may interact and react in a similar manner as a liquid mass does over a long period of time.

As to the origin of Kleopatra's companions, they could be a by-product of the spinning-up process leading to mass shedding in orbit. Such a scenario was already proposed to explain the binary nature of the asteroid 22 Kalliope (Descamps et al., 2008) and was recently substantiated by new spectroscopic observations of both components revealing the same surface composition (Laver et al., 2009). From Weidenschilling et al. (1989) we have computed the tidal evolution time scales as a function of the relative separation, and mass ratio. We may put corresponding values derived for the satellites of Kleopatra, $a/R_p$=10.04 and $q$=0.0003 for the outermost satellite and $a/R_p$=6.72 and $q$=0.00014 for the innermost satellite. The specific energy dissipation function $Q$ is estimated to ~100 by Yoder (1982). Recently, Goldreich and Sari (2009) demonstrated that a rubble pile is weaker than a monolithic body of the same composition. This represents a reduction in effective rigidity so that $\mu_{rubble} \sim (g\rho R\mu/\epsilon_Y)^{1/2}$, where $\epsilon_Y$ is the yield strain. Adopting values of $\epsilon_Y \sim 10^{-2}$, $\mu \approx 5 \times 10^{11}$ dyne cm$^{-2}$ for a stony-iron object, $\rho = 3.6$ g/cm$^3$, R = 67.5 km, we derive an effective rigidity lowered to $1.5 \times 10^9$ dyne cm$^{-2}$. Hence we have adopted $\mu Q \approx 1.5 \times 10^{11}$ dynes/cm$^2$. We can then bind the evolution time, assuming initial $a/R_p$=1, between ~100 million years for the outer satellite and between ~10 million years for the inner satellite. Thus their tidal evolution proceeds ~100 times faster than it would if they were monoliths.

---

[1] The normalized spin rate is defined by the ratio $\Omega=\omega_p/\omega_c$ where $\omega_p$ is the primary rotation rate and $\omega_c$ the critical spin rate for a spherical body.
[2] the non-sphericity parameter $\lambda$ is defined as the ratio between the moment of inertia of the body with respect to its spin axis, considered as the maximum moment of inertia axis, and the moment of inertia of the equivalent sphere ($2/5MR_e^2$).

## 5. Conclusion

New high resolution observations of asteroid (216) Kleopatra revealed two 5-10 km diameter moonlets whose orbits, after being fit by a Keplerian model, allowed us to derive the mass of Kleopatra. For the first time the real shape of this puzzling asteroid was directly and accurately imaged and compared to its reconstructed dog-bone radar shape model. For the most part the radar shape model is in good agreement with the observations but the AO observations revealed nevertheless a slightly more elongated body. Using AO imaging, thermal-infrared spectroscopy and past stellar occultations we were able to accurately determine the equivalent radius of Kleopatra which we found in agreement with its IRAS value ($R_{eq} = 67.5$ km). From this comprehensive work based on observations collected using a broad range of techniques, the average bulk density of this M-type asteroid was estimated to 3.6±0.4 g/cm$^3$. Kleopatra is therefore a loosely bound jumble of rock and metal chunks. This complex triple system could be a core remnant of a collisionally disrupted, differentiated asteroid. This catastrophic event could have happened more than 3 billion years ago.

**Acknowledgements**

These data were obtained with the W.M. Keck Observatory, which is operated by the California Institute of Technology, the University of California, Berkeley and the National Aeronautics and Space Administration. The observatory was made possible by the generous financial support of the W.M. Keck Foundation. The authors extend special thanks to those of Hawaiian ancestry on whose sacred mountain we are privileged to be guests. Without their generous hospitality, none of the observations presented would have been possible. FMA was supported by the National Science Foundation under award number AAG-0807468.

We would like to dedicate this work to our colleague Steve Ostro. Steve had a special relationship with the asteroid (216) Kleopatra, since he led the team which determined its unusual dog-boned shape using radar observations. Shortly after the publication of our IAU circular about these two moons, Steve wrote to us to share his enthusiasm about this discovery. Steve was a remarkable scientist, the founder of radar astronomy, but also a generous and caring person. We are grateful for the time he was here and for his legacy.

**Table 1:** Astrometric relative positions and estimated diameters of (216) Kleopatra's satellites were derived from AO images obtained with the 10-m Keck telescope. The diameter of each moonlet was estimated from peak to peak differential amplitude (Δmag) with the primary assuming similar albedos, which is sized to its IRAS diameter. Uncertainties in Δmag and satellite diameters are the standard deviations of the measurements.

| Date (UTC) | X (arcsec.) | Y (arcsec.) | Filter | Δmag | Diam (km) |
|---|---|---|---|---|---|
| S/2008 ((216)) 1 | | | | 5.9 ±0.4 | 8.9±1.6 |
| 2008- 9-19  6:17 | -0.25 | 0.50 | FeII | 6.05 | 8.3 |
| 2008- 9-19  8:44 | -0.34 | 0.54 | FeII | 6.01 | 8.5 |
| 2008- 9-19 11:38 | -0.44 | 0.57 | FeII | 5.5 | 10.6 |
| 2008- 9-19 11:51 | -0.44 | 0.57 | Jcont | 5.5 | 10.6 |
| 2008- 9-19 12: 2 | -0.44 | 0.58 | Kcont | 6.2 | 7.8 |
| 2008-10- 5  9:13 | -0.29 | 0.46 | Jcont | 6.3 | 7.5 |
| 2008-10- 5  9:49 | -0.31 | 0.47 | FeII | 5.3 | 11.4 |
| 2008-10- 5 10: 3 | -0.32 | 0.46 | H | 6.5 | 6.7 |
| 2008-10- 9  9:36 | 0.28 | -0.44 | FeII | 5.8 | 9.4 |
| S/2008 ((216)) 2 | | | | 6.5±0.4 | 6.9±1.6 |
| 2008- 9-19 11:38 | -0.18 | 0.35 | FeII | 6.4 | 6.9 |
| 2008- 9-19 11:51 | -0.20 | 0.36 | Jcont | 6.7 | 6.2 |
| 2008-10- 5  9:13 | -0.27 | 0.37 | Jcont | 6.8 | 5.9 |
| 2008-10- 5  9:49 | -0.29 | 0.39 | FeII | 5.7 | 9.9 |
| 2008-10- 5 10: 3 | -0.32 | 0.39 | H | 6.9 | 5.6 |
| 2008-10- 9  5:46 | -0.32 | 0.29 | FeII | 6.3 | 7.4 |

**Table 2:**
Orbit elements in mean J2000 equator and equinox and derived parameters. Uncertainties are formal and subjected to the assumption of zero eccentricity. The reference epoch is Julian date 2454728.5 (2008.715947981).

| Fitted orbit parameters | | S/2008 ((216)) 2 | S/2008 ((216)) 1 |
| --- | --- | --- | --- |
| Period (days) | $P$ | 1.24 ± 0.02 | 2.32 ± 0.02 |
| Semi-major axis (km) | $a$ | 454 ± 6 | 678 ± 13 |
| Inclination (°) | $i$ | 49 ± 2.0 ° | 51 ± 2.0 ° |
| Longitude of Ascending node (°) | | 160 ± 3.0° | 166 ± 3.0° |
| Derived Parameters | | | |
| System Mass (x$10^{18}$ Kg) | | 4.67 ± 0.2 | 4.63 ± 0.2 |
| Orbit pole right ascension(°) | $\lambda$ | 74 ± 2° | 79 ± 2° |
| Orbit pole declination (°) | $\beta$ | 16 ± 1° | 16 ± 1° |

**Table 3:** Longitude and Latitude of the Kleopatra sub-Earth point (SEP) for each AO observation. Phase angle and geocentric distance are also given.

| Date (UTC) | Lon SEP (°) | Lat SEP (°) | Phase angle (°) | Geocentric distance (ua) |
|---|---|---|---|---|
| 2008- 9-19  6:17 | 152.1 | -9.3 | 7.5 | 1.24 |
| 2008- 9-19  8:44 | 347.1 | -9.3 | 7.5 | 1.24 |
| 2008- 9-19 11:38 | 153.3 | -9.2 | 7.5 | 1.24 |
| 2008- 9-19 11:51 | 138.9 | -9.2 | 7.5 | 1.24 |
| 2008- 9-19 12: 2 | 126.6 | -9.2 | 7.5 | 1.24 |
| 2008-10- 5  9:13 | 206.3 | -5.7 | 12.3 | 1.25 |
| 2008-10- 5  9:49 | 166.2 | -5.7 | 12.3 | 1.26 |
| 2008-10- 5 10: 3 | 150.6 | -5.7 | 12.3 | 1.26 |
| 2008-10- 6 9:49 | 1.96 | -5.5 | 12.3 | 1.26 |
| 2008-10- 9 9:36 | 243.8 | -4.9 | 13.8 | 1.27 |
| 2008-10- 9 5:46 | 139.9 | -5.0 | 13.8 | 1.27 |

**Figure 1**: The triple system (216) Kleopatra as it was discovered with the Keck II AO system and NIRC2 camera at ~1.2 μm. The left panel shows the image recorded on September 19, 2008 at 11:51 UTC in Jcont filter ($\lambda_c$ = 1.2132, $\Delta\lambda$=0.0198 μm) while the right panel shows the image obtained after deconvolution and applying Laplacian filtering. North is up and East is left. In the upper rightmost part of the images two tiny moons, lined up with Kleopatra, begin to emerge from the glare of the asteroid. The deconvolved image highlights Kleopatra's odd shape.

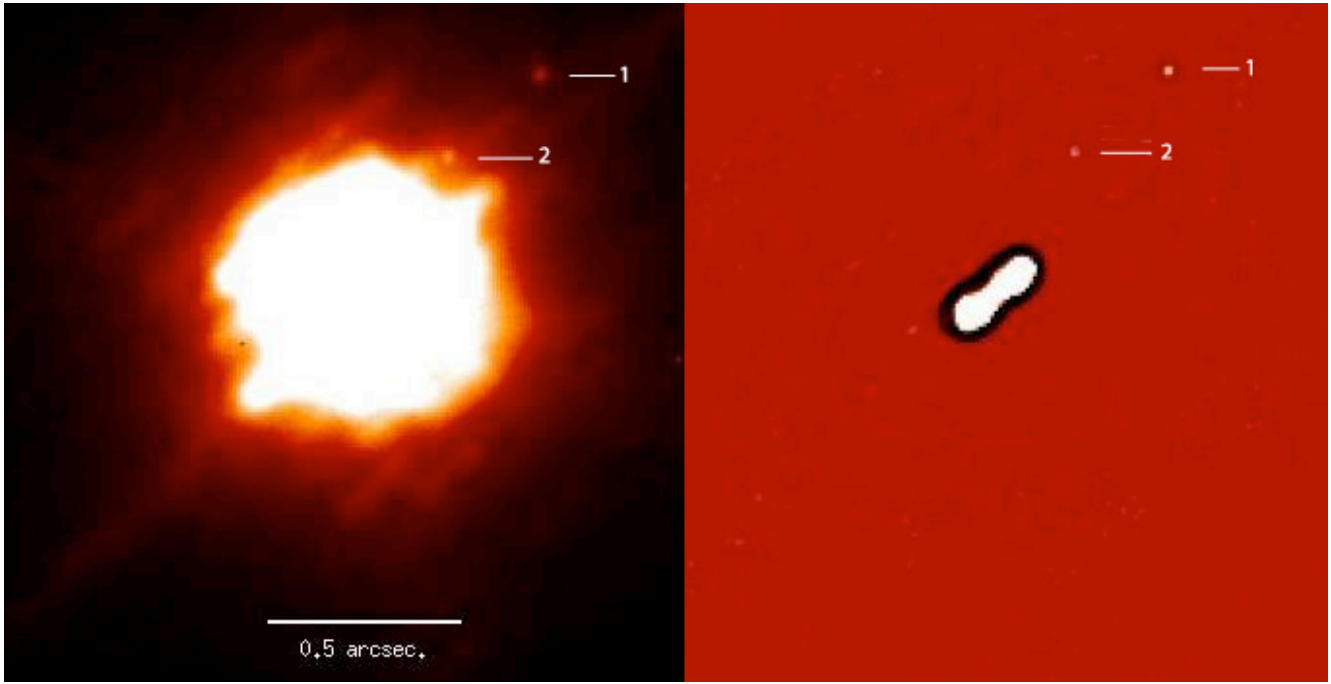

**Figure 2**: Apparent orbits of the two known satellites of Kleopatra. Astrometric positions have been reported along with their residuals between measured positions (cross) and computed positions (dots).

**Figure** 3: A few lightcurves of Kleopatra recorded in 2007 and 2008. Synthetic lightcurves (solid lines) derived from the radar model are superimposed to the observed data (cross).

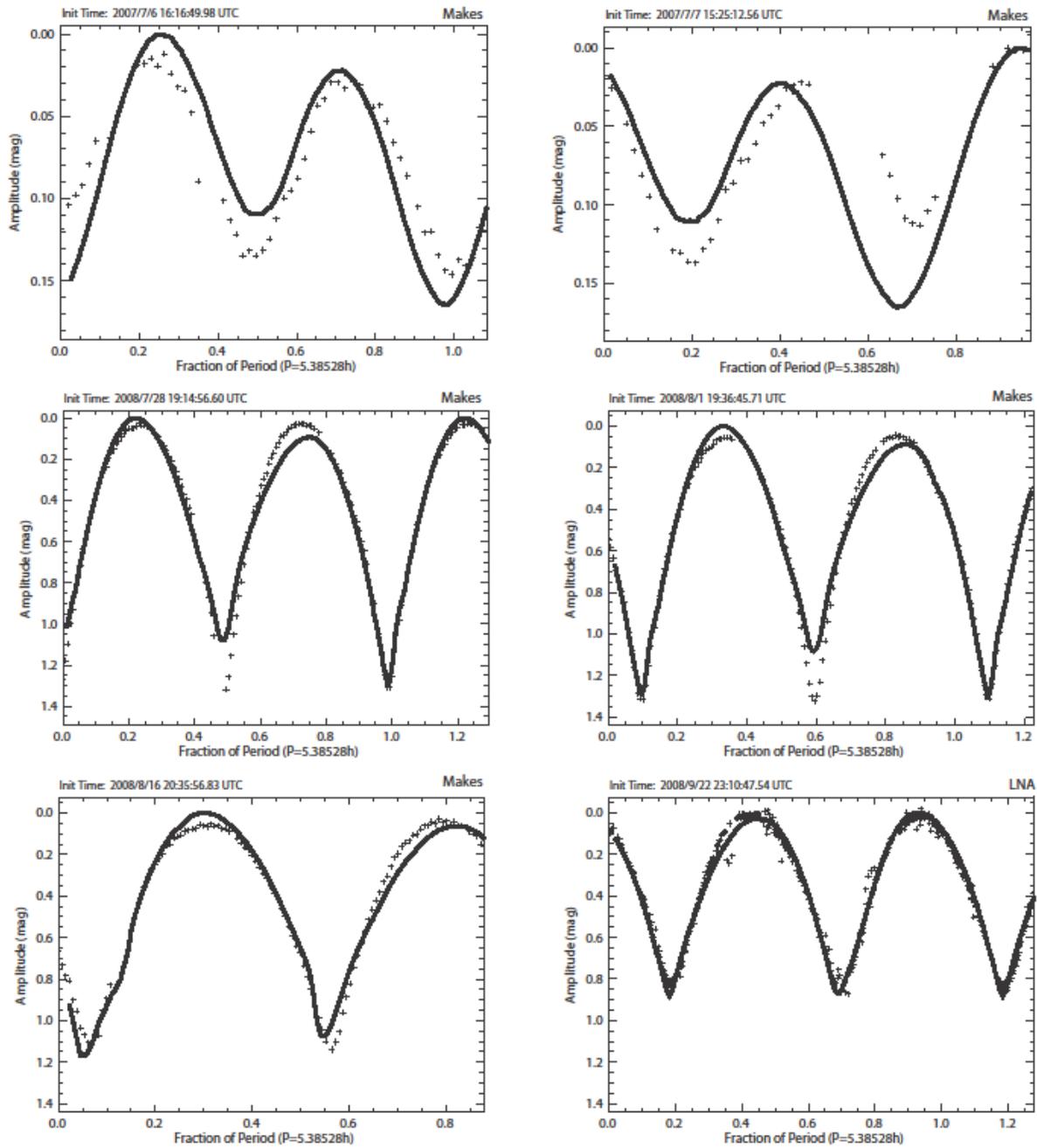

**Figure** 3: Lightcurves of Kleopatra – continued

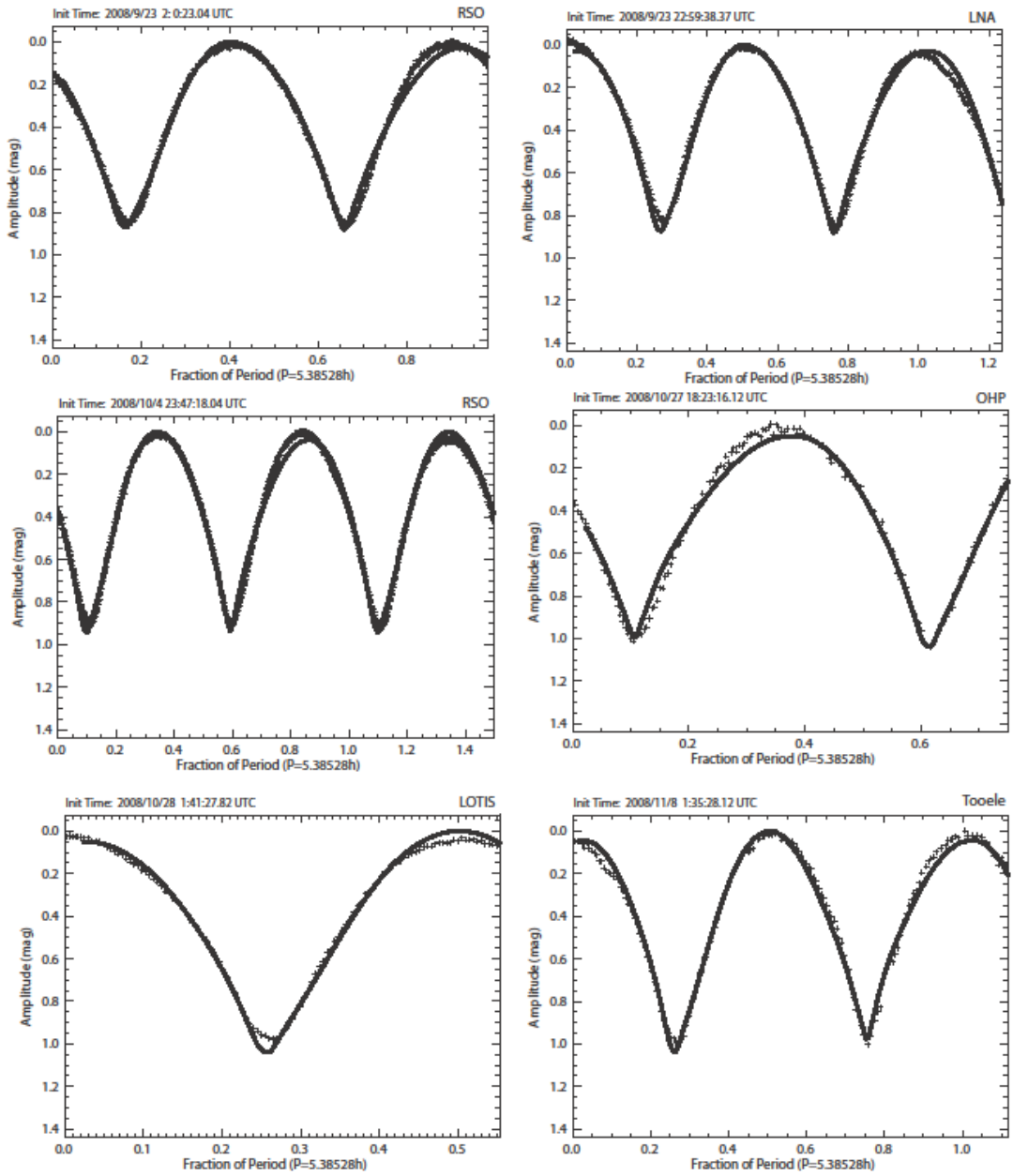

**Figure** 3: Lightcurves of Kleopatra – continued

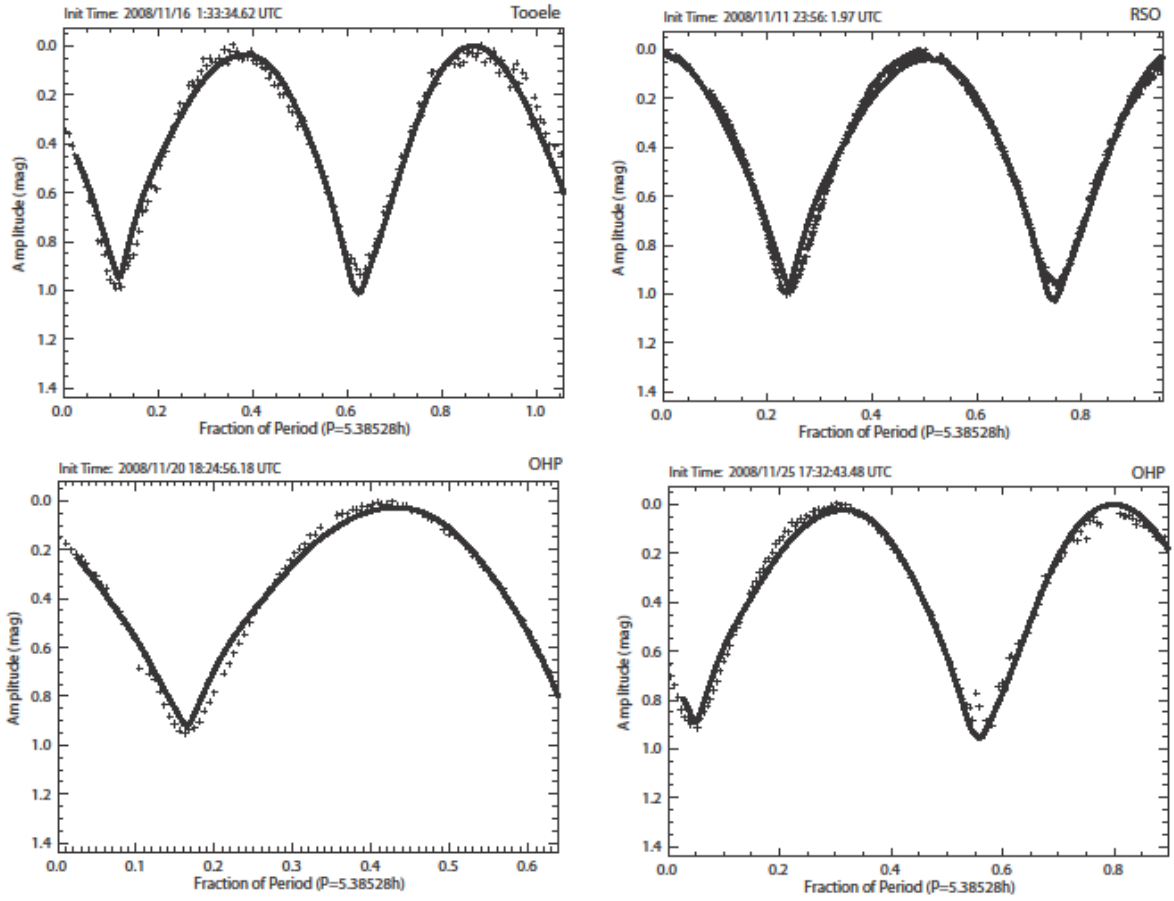

**Figure 4:** Adaptive optics images of Kleopatra taken with the 10-m Keck telescope in September and October 2008. A Laplacian filter has been applied to the deconvolved images in order to enhance the overall contour of Kleopatra. The radar non-convex shape solution is shown for the corresponding observing dates. North is up and East is left. The spatial resolution is 35 mas (28 km). Contour lines of the radar shape model are superimposed to the observations for two values of the equivalent radius, 67.5 km (solid line, IRAS radius) and 54.25 km (dotted line, radar radius (Ostro, 2000)).

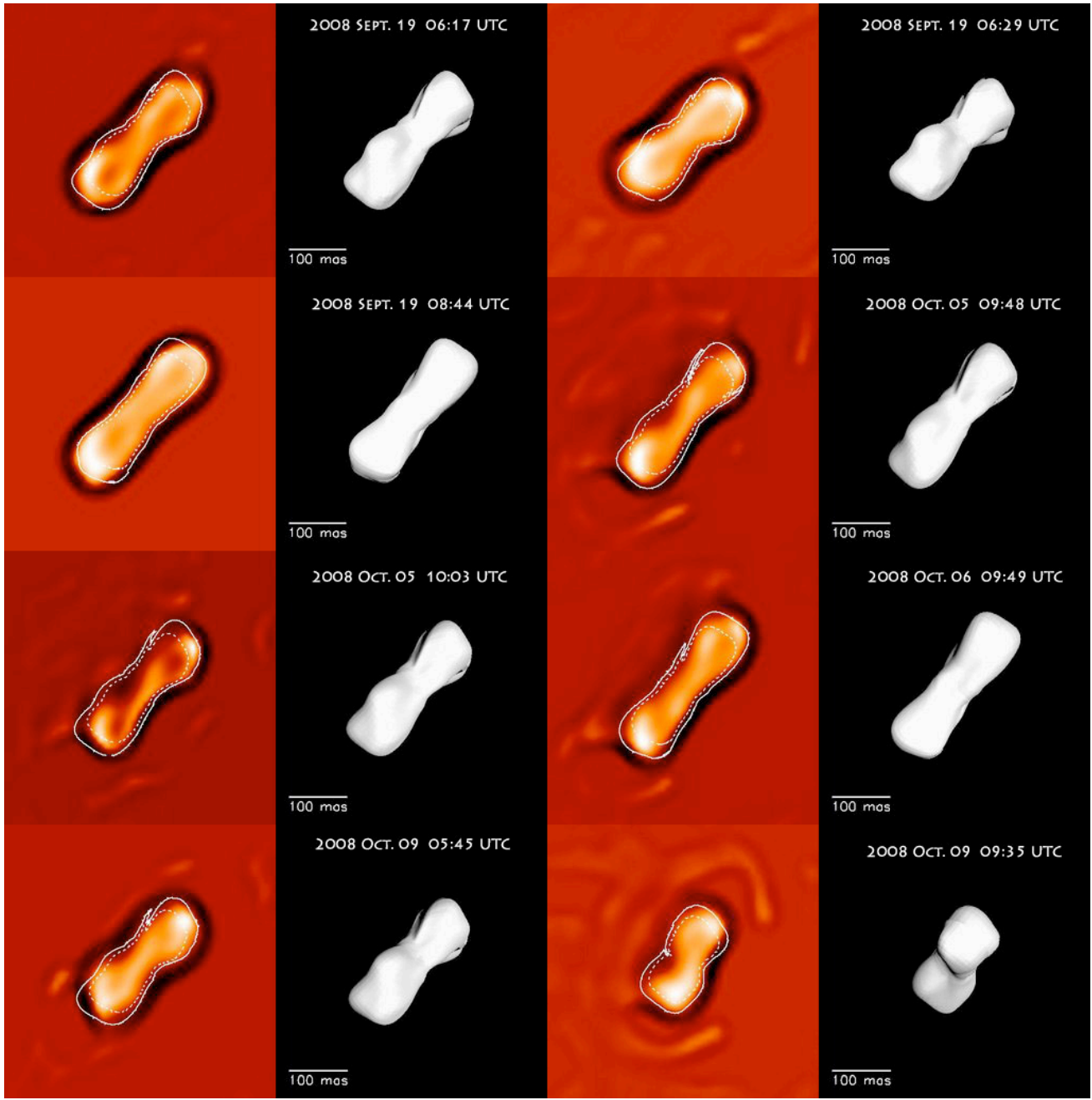

**Figure** 5: Thermal flux and high resolution spectra of (216) Kleopatra recorded using Spitzer/IRS on February 2, 2006 from 06:07 to 06:21 UT. We derived an effective radius of 78.2 ± 2.9 km, a visible geometric albedo of 0.087 -/+ 0.007 and a beaming factor of 1.25 ± 0.03 after fitting with a version of the NEATM model that was modified to incorporate ellipsoidal shapes. An inset shows the appearance of the primary (FOV of 0.18" x 0.18") based on the radar shape model as seen from the Spitzer telescope at the time of the observation. The same effective radius given by the cross-section image is obtained for an equivalent radius of (216) Kleopatra of 67.5 km.

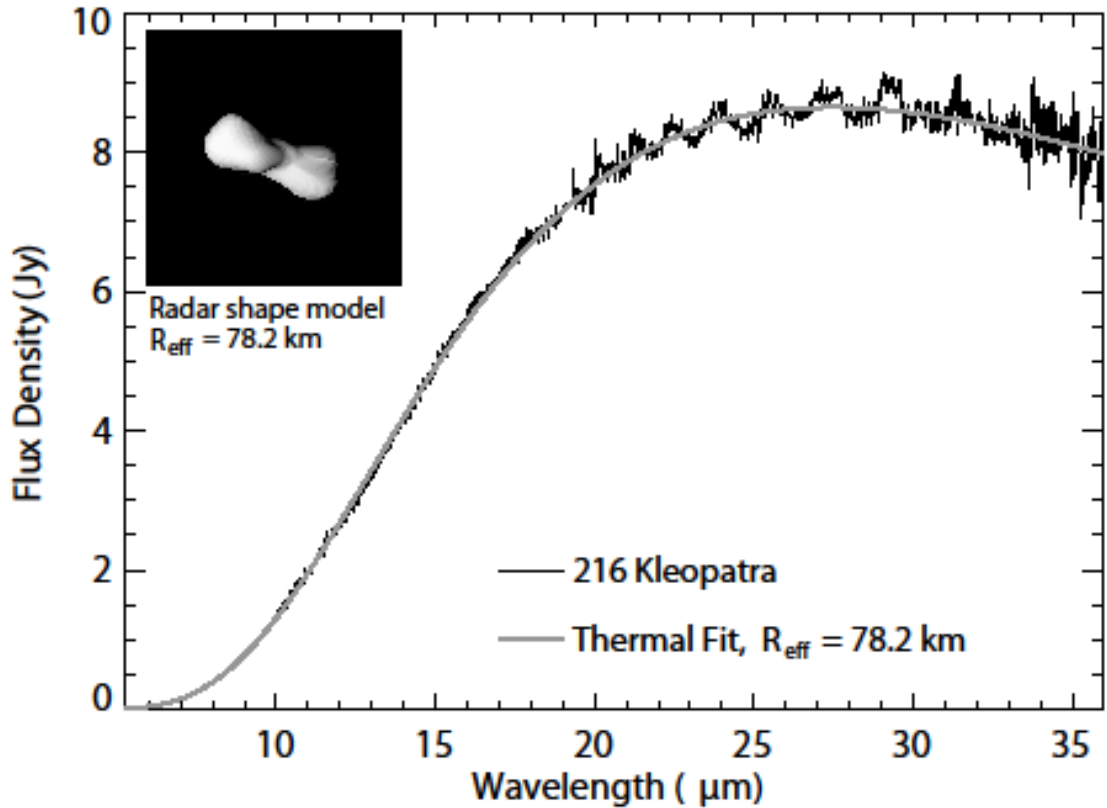

**Figure** 6: Stellar occultation of 1980 Oct. 10. Overall view of the occultation: observed chords (plain lines), absence of detection (dashed lines), Kleopatra's silhouette computed by our model, and the orbits of the two satellites based on the orbital parameters of this work (dots).

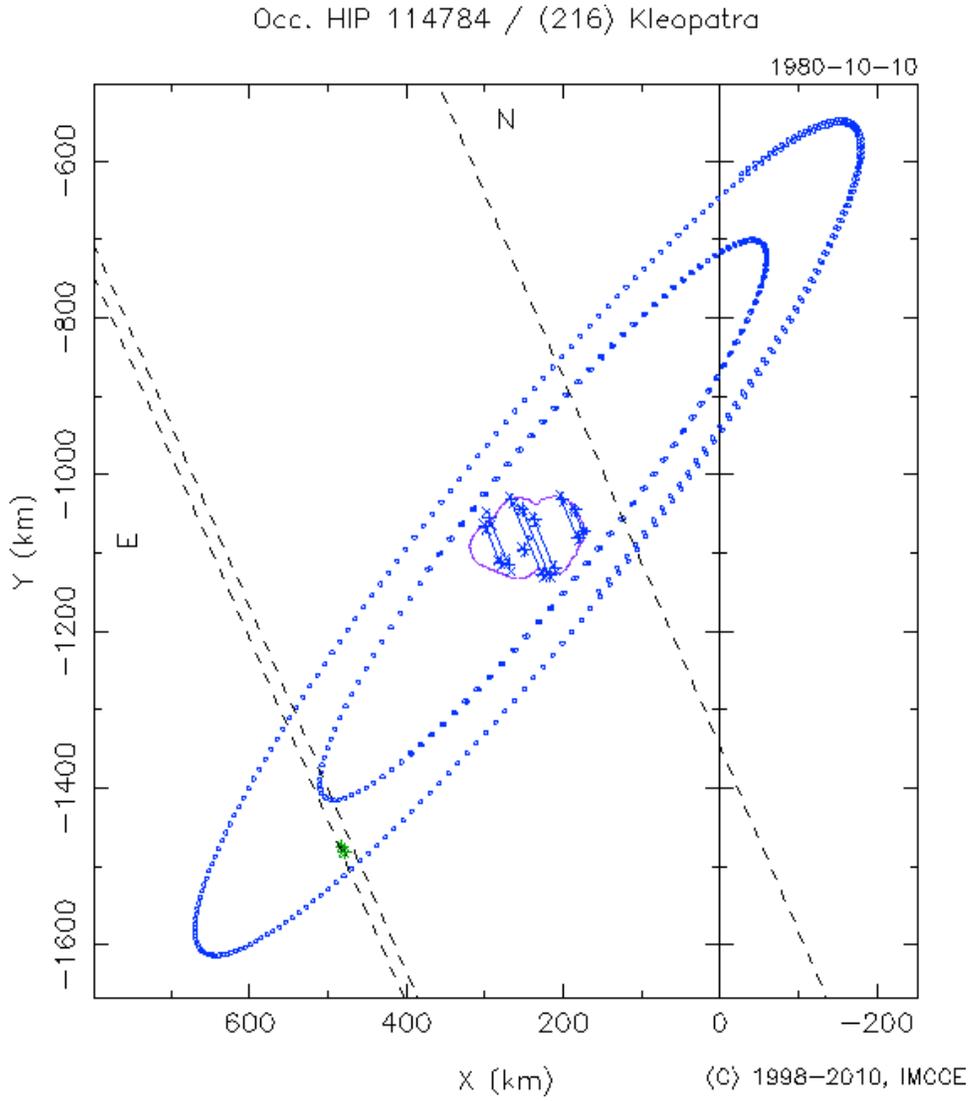

**Figure** 7: Stellar occultation of 1980 Oct. 10. Close-up view of the observed chords superimposed to the computed profile of Kleopatra obtained from the radar shape model.

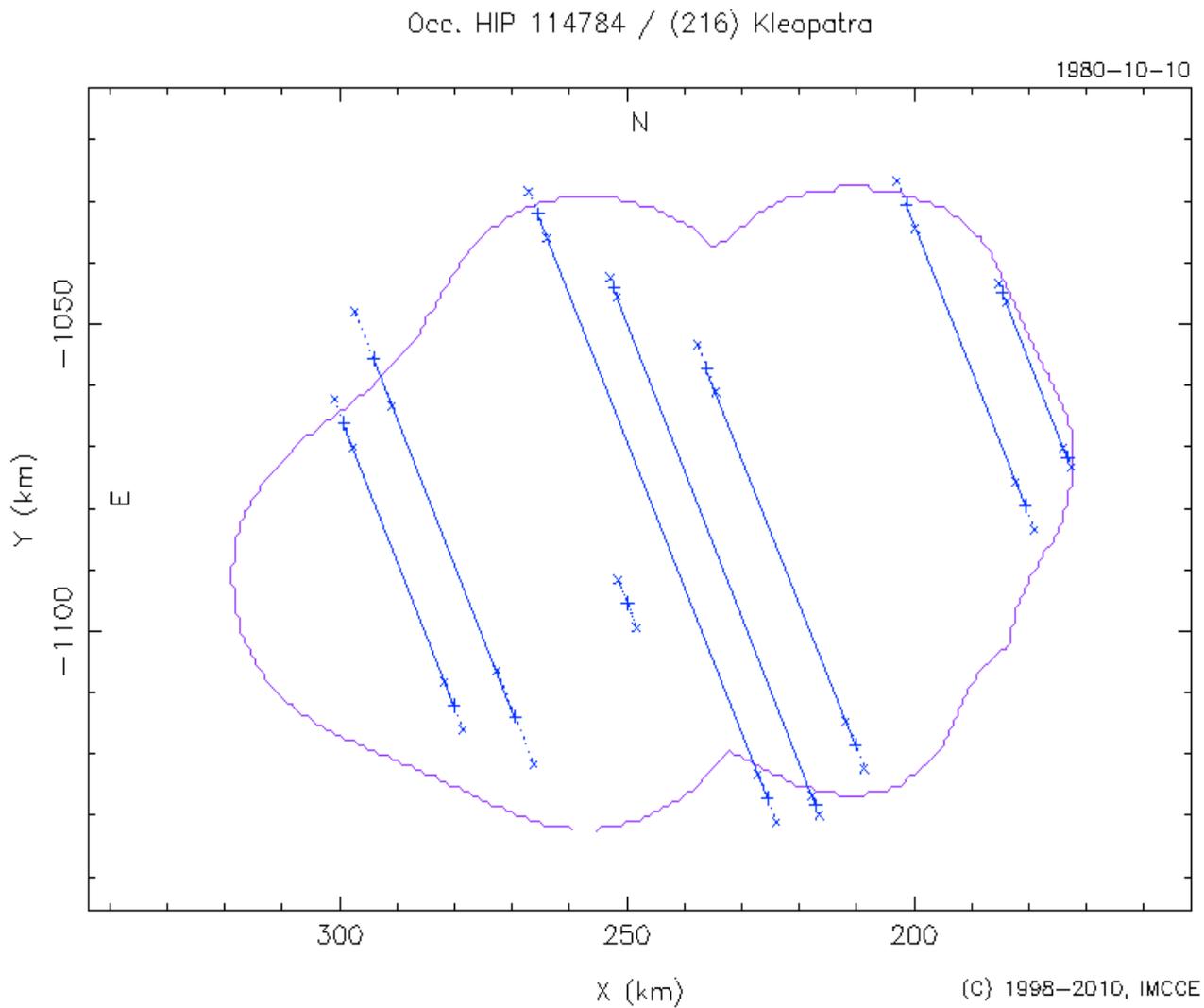

**Figure** 8: Stellar occultation of 1980 Oct. 10. Close-up views of the observed chord of the secondary event and the computed orbits of Kleopatra's moonlets S/2008 ((216)) (dots).

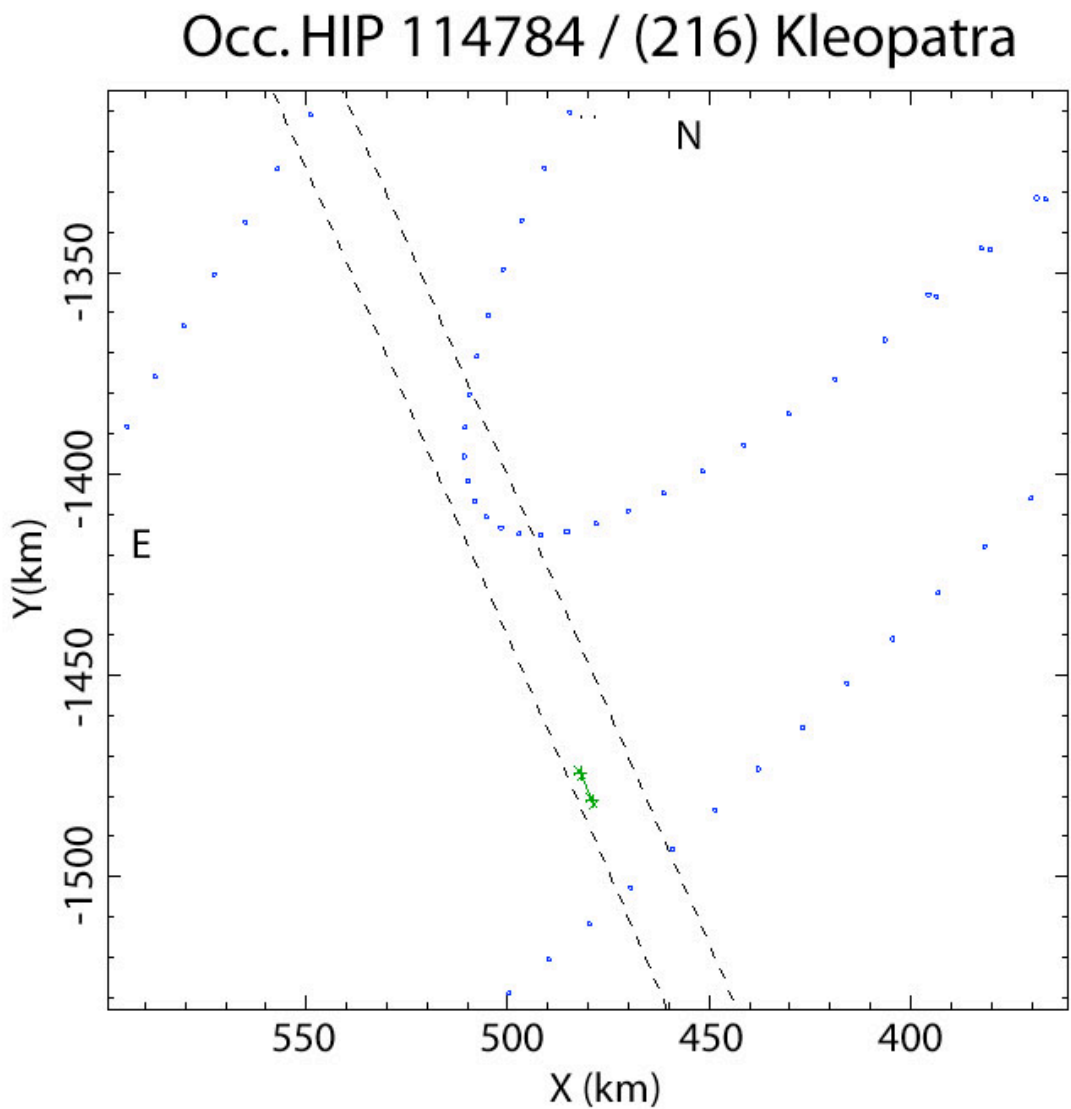

**Figure** 9: Stellar occultation of 1991 Jan. 19. The solid line denotes the contour of Kleopatra inferred from the AO observation made on 2008 October 5 10:03 UTC. At that time the central meridian longitude was 150.6°. The dashed line shows the contour line based on the AO observation made on 2008 September 19 6:17 UTC. In 1991 the central meridian longitude was 152.6° with a north pole position angle of 322° whereas it was of 52° in 2008 so that a rotation of 270° of the contour profiles allows us to superimpose them to the picture of occultation chords. The discrepancies mainly result from the central meridian latitude which is shifted by 30° between both epochs.

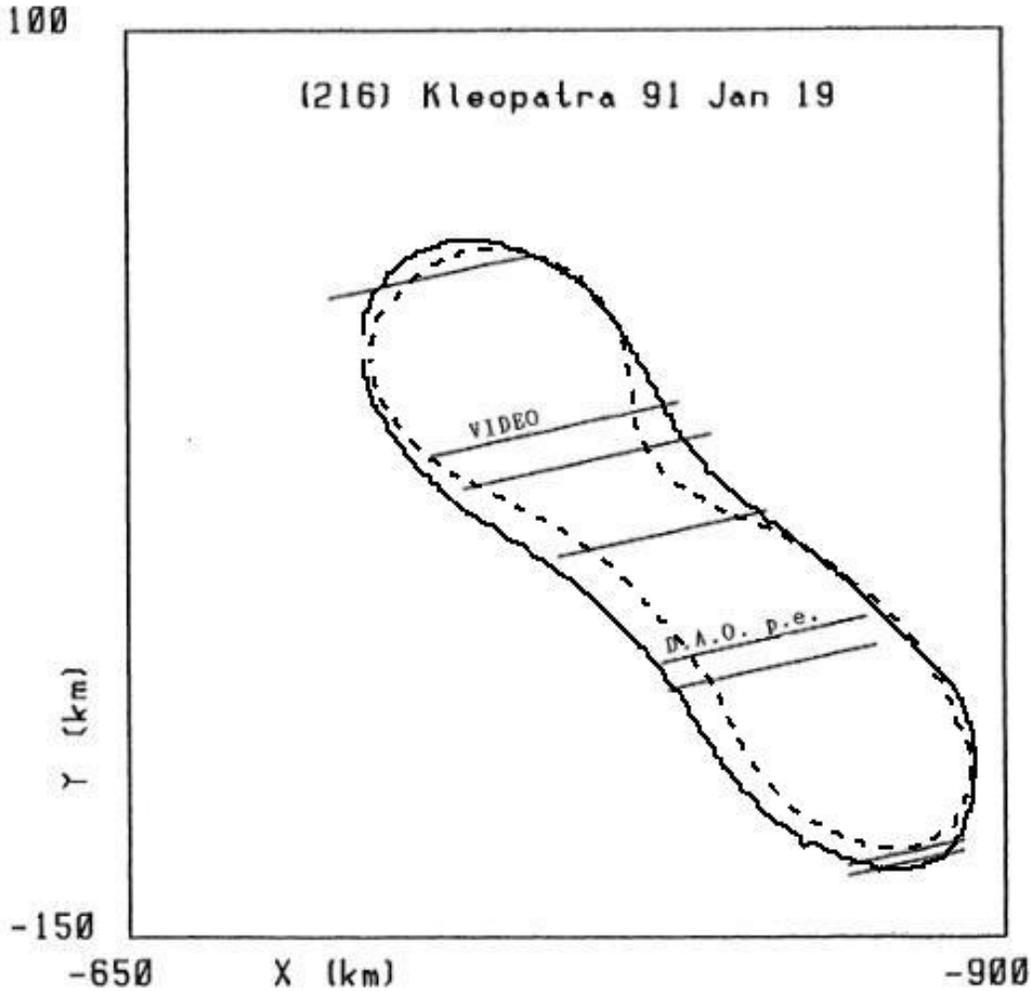